# Barocaloric thermal batteries

Zhe Zhang[1,2,#], Kuo Li[3,#], Shangchao Lin[4,#,*], Ruiqi Song[1], Dehong Yu[5], Yida Wang[3], Jingfan Wang[6], Shogo Kawaguchi[7], Zhao Zhang[1,2], Chenyang Yu[1,2], Xiaodong Li[8], Jie Chen[10], Lunhua He[9,10,11], Richard Mole[5], Bao Yuan[8,10], Qingyong Ren[8,10], Kun Qian[4], Zhuangli Cai[4], Jingui Yu[12], Mingchao Wang[13], Changying Zhao[4], Xin Tong[8,10], Zhidong Zhang[1,2,*], and Bing Li[1,2,*]

[1]Shenyang National Laboratory for Materials Science, Institute of Metal Research, Chinese Academy of Sciences, 72 Wenhua Road, Shenyang, Liaoning 110016, China.
[2]School of Materials Science and Engineering, University of Science and Technology of China, 72 Wenhua Road, Shenyang, Liaoning 110016, China.
[3]Center for High Pressure Science and Technology Advanced Research, Beijing, China.
[4]Key Laboratory for Power Machinery and Engineering of Ministry of Education, Institute of Engineering Thermophysics, School of Mechanical and Power Engineering, Shanghai Jiao Tong University, Shanghai, 200240, China.
[5]Australian Nuclear Science and Technology Organisation, Locked Bag 2001, Kirrawee DC NSW 2232, Australia.
[6]Department of Mechanical Engineering, Florida State University, Tallahassee, Florida 32310, United States.
[7]Japan Synchrotron Radiation Research Institute, SPring-8, 1-1-1 Kouto, Sayo-cho, Sayo-gun, Hyogo 679-5198, Japan.
[8]Institute of High Energy Physics, Chinese Academy of Sciences, Beijing 100049, China.
[9]Beijing National Laboratory for Condensed Matter Physics, Institute of Physics, Chinese Academy of Sciences, Beijing 100190, China.
[10]Spallation Neutron Source Science Center, Dongguan 523803, China
[11]Songshan Lake Materials Laboratory, Dongguan 523808, China.
[12]School of Mechanical and Electronic Engineering, Wuhan University of Technology, Wuhan, 430070, China.
[13]Centre for Theoretical and Computational Molecular Science, Australian Institute for Bioengineering and Nanotechnology, The University of Queensland, St. Lucia, QLD 4072, Australia.
* Corresponding authors: bingli@imr.ac.cn, zdzhang@imr.ac.cn, shangchaolin@sjtu.edu.cn.
# These authors equally contributed to the research.

**Nowadays the world is facing a prominent paradox regarding thermal energy. The production of heat accounts for more than 50% of global final energy consumption[1] while the waste heat potential analysis reveals that 72% of the global primary energy consumption is lost after conversion mainly in the form of heat[2]. Towards global decarbonization, it is of vital importance to establish a solution to thermal energy utilization under full control[3-6]. Here, we propose and realize an unprecedented concept --- barocaloric thermal batteries based on the inverse colossal barocaloric effect of $NH_4SCN$. Thermal charging is initialized upon pressurization through an order-to-disorder phase transition below 364 K and in turn the discharging of 43 J g$^{-1}$, which are eleven times more than the input mechanical energy, occurs on demand at depressurization at lower temperatures. The discharging is also manifested as a directly measured temperature rise of 12 K. The thermodynamic equilibrium nature of the pressure-restrained heat-carrying phase guarantees stable storage and/or transport over a variety of temporal and/or spatial scales. The barocaloric thermal batteries reinforced by their solid microscopic mechanism are expected to significantly advance the ability to take advantage of waste heat.**

Even if roughly 90% of the world's energy use today involves heat over a wide range of temperatures, the current ability to manipulate heat is still limited[3-6]. The preferred strategy for thermal energy recycling is that waste heat is stored after harvesting and reused for heating purposes without further conversion. Towards such a goal, great obstacles have to be overcome to improve the controllability, tunability, and transportability of thermal energy. Thermal energy storage often employs phase-change materials thanks to their significant latent heat at phase transitions[7]. A schematic diagram is given in **Fig. 1a**. During their working processes, the temperature is the only dominating factor as they absorb heat from the environment when the temperature becomes higher than the phase transition temperature ($T_t$) and release heat to the environment vice versa. Given that the storage just takes effect at phase transition temperatures, a material selectivity is imposed and leads to limited applicability. To adapt to variable applications, multiple phase-change materials have to be taken into account[8]. The cost would be increased by such complexity. Differing from the electrical energy, long-distance transmission of heat is almost impossible due to the heat dissipation into the environment[4]. As a result, most harvested heat is wasted during transport.

Inspired by the recent discovery of colossal barocaloric effects (CBCEs) in plastic crystals[9], we propose the barocaloric thermal battery that is a brand-new concept based on inverse barocaloric effects. Plastic crystals are also known as orientation-disordered crystals where huge entropy changes are obtained at the order-to-disorder transitions[10]. As the plastic transitions are effectively tuned by external pressures, they are promising candidate working materials for solid-state refrigeration[9]. Unlike the normal barocaloric materials that release heat at pressurization, a material with an inverse barocaloric effect absorbs heat, in analogy to the pressure-induced ice melting[11]. This uniqueness allows us to design a barocaloric thermal battery, as schematically shown in **Fig. 1b**. The material is thermally charged through a pressure-induced order-to-disorder transition. At the same time, the heat sink contacting the material is cooled down. As long as the pressure is held, the material is equilibrated at the disordered state for heat storage, for which the long-distance transmission of heat becomes feasible. This equilibrium-state feature differs from the intrinsic metastability of other heat-storage materials[12,13]. The stored heat can be readily released at the transition back to the ordered state at depressurization and then recharging can be followed. In the present study, we materialize this concept in $NH_4SCN$ that is the first compound exhibiting an inverse CBCE. Taking advantage of neutron scattering and complementary multiscale simulations, we have established the mechanism in depth. As an emergent solution to manipulate heat, barocaloric thermal batteries are expected to play an active role in plenty of applications such as industrial waste heat harvesting, heat transfer systems of solid-state refrigeration, smart grid, and residential heat management.

The actual working processes of barocaloric thermal batteries are demonstrated in $NH_4SCN$. As shown in **Fig. 1c**, the temperature-pressure cycle is depicted by the dash lines on the phase diagram. The corresponding heat flow data are plotted in **Fig. 1d**. At 349 K the compound is thermally charged upon the pressure changing from 8.2 to 90 MPa, appearing as an endothermic peak of the heat flow as a function of time. The latent heat that is related to this peak is estimated to be 43 J g$^{-1}$. To mimic the storage/transport process, subsequently, the compound is cooled down to 313 K under 90 MPa. In turn, an exothermic peak is observed upon pressure decreased down to 8.2 MPa. The latent heat at discharging is well consistent with that at charging. Thermal charging and discharging processes have also been repeated at different temperatures as included in **Extended Data Fig. 1a** and **b**. They are also well reproduced using the finite-element (FE) thermal simulations. It can be seen in **Fig. 1d** that the simulated heat flow curve is in great agreement with the experiments. The performances of barocaloric thermal batteries of NH$_4$SCN are further exploited as a function of charging pressure and discharging temperature. It is no doubt that the material can be fully charged once the pressure is higher than the critical pressure (**Extended Data Fig. 1c**). As for the working temperature region, it can be seen in **Extended Data Fig. 1d** that the discharging can take place at 273 K when the charging pressure is 400 MPa.

The practical performances of barocaloric thermal batteries are further exploited at room temperature through the direct measurements on the temperature changes in response to the applied pressures up to 300 MPa. **Fig. 2** shows the pressure-induced temperature changes ($\Delta T$) at different pressure ramping rates. It can be seen that there is a pair of strong peaks, which are attributed to the charging and discharging of NH$_4$SCN. At the same time, there is another pair of much weaker peaks, corresponding to the mechanical works associated with the external pressures. As a hallmark for an inverse barocaloric effect, note that these two pairs of peaks have different signs and these different signals are partially canceled. It can be seen that $\Delta T$ is proportional to the pressure ramping rate. At the fastest experimental rate in our measurements, as shown in **Fig. 2c**, $\Delta T$ at discharging is as high as 12 K. This pressure ramping rate dependence reflects the differences of adiabatic conditions in the charging-discharging processes. As such, the simulated process with a significantly short thermal exchange period indicates that theoretical $\Delta T$ is about 26 K (see the inset of **Fig. 2c**). The complete process is also recorded in **Supplementary Video 1**. As far as a real application is concerned, the ability to release heat under full control is highly desirable.

As the concept of barocaloric thermal batteries is proposed and realized with excellent performances, it is crucial to establish the physical mechanism for further improvement towards large-scale practical applications. What makes the barocaloric thermal batteries feasible is the inverse CBCE of NH$_4$SCN. This compound has unique barocaloric performances not only because it is the first example bearing an inverse CBCE, but also because its colossal entropy changes accommodate the colossal relative cooling power (RCP). NH$_4$SCN undergoes a phase transition at about 364 K on heating while 335 K on cooling under ambient pressure. The temperature dependencies of entropy changes at different pressure changes ($\Delta S_{P_0 \to P}$) are plotted in **Fig. 3a**. The maximum entropy changes are ($\Delta S_{P_0 \to P}^{max}$) about 128.7 J kg$^{-1}$ K$^{-1}$ and are one order of magnitude larger than those of other current barocaloric materials (see **Fig. 3b**), after which the inverse CBCE is named. The entropy changes at varying pressures indicate that the saturation pressure is about 40 MPa, in contrast to other systems whose saturation pressures are usually higher than 500 MPa [14-19]. This leads to superior normalized entropy changes, $|\Delta S_{P_0 \to P}^{max}/P| = 1.61$ J kg$^{-1}$ K$^{-1}$ MPa$^{-1}$. The phase transition temperature ($T_t$) is shifted to the lower temperatures at larger pressures. The slope defines the phase boundaries on the *T-P* phase diagram shown in **Fig. 1c**[20,21]. At the pressure change from 0.1 to 100 MPa, $T_t$ (on heating) is shifted from 364 K down to 334 K, giving rise to $|dT_t/dP|$ of about 0.3 K MPa$^{-1}$, which is also the largest as compared to other leading materials in **Fig. 3c**. For this reason, the obtained RCP (evaluated as the integrated area under entropy change curves[22]) is the largest as well (**Fig. 3d**). Compared to the previous reports[16] where a smaller $|dT_t/dP|$ leads to a larger entropy change and yet to a smaller RCP, the colossal entropy changes and RCP are simultaneously achieved in NH$_4$SCN (see **Extended Data Fig. 2**), which is benefited from the huge lattice volume changes to be described below.

Previous investigations suggest that NH$_4$SCN undergoes successive phase transitions from monoclinic to orthorhombic, and further to tetragonal phases at about 360 and 390 K, respectively[23-26]. The inverse CBCE is attributed to the former. Our neutron diffraction measurements on deuterated samples (ND$_4$SCN) confirm the crystal structures of the three phases. The information of the deuterated samples is summarized in **Extended Data Fig. 3**. Shown in **Fig. 4a** and **b** are the crystal structures and refinements of the diffraction patterns for the monoclinic and orthorhombic phases, respectively. In the monoclinic phase, ND$_4^+$ and SCN$^-$ are both ordered. In particular, SCN$^-$ ions are oriented in an antiparallel fashion. As for the orthorhombic phase, D atoms in ND$_4^+$ ions become disordered, whose trajectories form a simple cubic lattice. Given that the configuration of ND$_4^+$ is rigid, there are six possible distinguishable orientations for the ND$_4^+$ tetrahedron in this disordered motif. If two D atoms occupy two sites of a face diagonal, another two D atoms reside on the two sites of the perpendicular face diagonal on the opposite plane. Currently, it is still unclear whether SCN$^-$ ions are ordered in the orthorhombic phase. It is also difficult to distinguish the exact arrangement in our diffraction data. Since our molecular dynamics (MD) simulations favor the ordered state, we adopt this ordered model in our refinement. The tetragonal phase is characteristic of the disordered ND$_4^+$ and SCN$^-$ ions as plotted in **Extended Data Fig. 4a**. The detailed crystallographic data of the three phases are summarized in **Extended Data Table 1**.

Focusing on the monoclinic-to-orthorhombic phase transition, the synchrotron X-ray diffraction measurements are conducted at a selected series of temperatures. Based on the structural models derived in the refinements of neutron diffraction data, we calculate the lattice dimensions as a function of temperature. With warming up from room temperature, the phase transition takes place between 355 and 365 K. Two-phase coexistence is observed at 360 K. At the phase transition, the unit cell volume exhibits an abrupt contraction by about 5% as shown in **Fig. 4c**, which corresponds to a giant specific volume change of about -3.8×10$^{-5}$ m$^3$ kg$^{-1}$. To our knowledge, this negative thermal expansion is the highest value among known caloric materials. This is just responsible for the combination of large $\Delta S_{P_0 \to P}^{max}$ and RCP because the huge volumetric change cancels the negative contribution of $|dT_t/dP|$ to entropy changes. In the orthorhombic phase, the SCN⁻ ions are confined in the *ab* plane. At the orthorhombic-to-monoclinic transition, the SCN⁻ ions tilt out of the *ab* plane to the *ac* plane, which leads to the increase of lattice constant *a* and *c*, corresponding to *b* and 2*a* in the monoclinic setting (see **Extended Data Fig. 4e**). Such a negative thermal expansion is also confirmed in the predicted lattice constants of the monoclinic and orthorhombic phases from both first-principle density functional theory (DFT) and MD simulations (**Extended Data Table 2**).

Further, both the reorientation dynamics and collective dynamics (phonons) are explored using quasi-elastic neutron scattering (QENS) and inelastic neutron scattering (INS), respectively. Since hydrogen atoms have an enormous incoherent scattering length, the reorientation dynamics of $NH_4^+$ can be well probed by QENS[27]. The QENS spectra as a function of energy transfer (*E*) at different momentum transfer (*Q*) are plotted in **Extended Data Fig. 5a**, **b,** and **c** at 300, 380, and 400 K, corresponding to the three phases, respectively. The broadening of the elastic line signals the QENS. It can be seen that the QENS intensity becomes stronger as *Q* increases. At a specific *Q* position, the intensities at 380 and 400 K are almost identical, which are much stronger than that of 300 K. Therefore, the monoclinic-to-orthorhombic phase transition is accompanied by modifications of the reorientation dynamics while the orthorhombic-to-tetragonal one is not.

Reorientation dynamics are revealed using spectral fitting. The QENS spectra are fitted to a combination of a delta function and a Lorentzian function plus a linear background, which are convoluted to the instrumental resolution function. An example is given at $Q = 1.2$ Å⁻¹ in **Extended Data Fig. 5d**, **e,** and **f**. The integrated area of the delta and Lorentzian components represent the elastic and QENS intensities, respectively. Hence, elastic incoherent scattering factor (EISF) is defined as the ratio between elastic intensity and the total intensity to evaluate the reorientation dynamics. Shown in **Fig. 4d** and **e** are EISF as a function of *Q* for $NH_4SCN$ at 300 K for the monoclinic phase and at 380

K for the orthorhombic phase, respectively. The experimental data are compared with a few models for $NH_4^+$ ions including two-fold ($C_2$) or three-fold ($C_3$) rotation, tetrahedral tumbling, cubic tumbling, and isotropic rotation, whose details[28] are given in **Methods**. It can be seen that the 300 K data are close to the tetrahedral tumbling model, where the $NH_4^+$ tetrahedron is as a whole randomly oriented over four indistinguishable configurations. Even if the high-$Q$ data were not accessed due to the limited $Q$ region of the instrument, this mode is supported by the crystal structure data shown in **Fig. 4a** as well as the MD-predicted orientational distribution contour plots (inset of **Fig. 4d**) and dynamic trajectories (**Extended Data Fig. 6c and e**) of the N-H bonds of $NH_4^+$ in the monoclinic phase.

Nevertheless, the reorientation dynamics become more complicated at 380 K. As shown in **Fig. 4e**, the data points are most likely described by the cubic tumbling mode even if a sizable deviation is present. Recalling the crystal structure shown in **Fig. 4b**, the orientation of the tetrahedron of $NH_4^+$ is random over a cubic lattice, which resembles the cubic tumbling model. However, it is clear that the experiment data points deviate from this model. The MD simulations suggest that the distribution is not a simple cubic lattice, but with preference (see the inset of **Fig. 4e**). More careful analysis of the crystal structure indicates that the hydrogen-bonding environment of $NH_4^+$ is not exactly symmetric. MD-predicted orientational surface plots (see inset of **Fig. 4e**) and dynamic trajectories of the N-H bonds of $NH_4^+$ also suggest that such a cubic tumbling mode shows spatial preference (**Extended Data Fig. 6a, b,** and **d**). At 400 K, the system becomes tetragonal. As mentioned before, the QENS spectra of the orthorhombic phase and the tetragonal phase are almost identical, which suggests that the motions of $NH_4^+$ are the same.

The average timescales ($\tau$) of the reorientation modes are roughly characterized by the full width at half maximum ($\Gamma$) of the Lorentzian component using $\tau = 2\hbar/\Gamma$[29]. In **Fig. 4f**, the $Q$ dependence of $\tau$ is summarized at 300 and 380 K for the monoclinic and orthorhombic phases, respectively. The weak $Q$ dependence indicates the local nature of the modes. A weak $Q$ dependence has also been found in a similar system of NaAlH$_4$ (ref. 30). It can be seen that the preferred cubic tumbling mode is much faster than the tetrahedral tumbling. $\tau$ can be also estimated using MD simulations for each phase as shown in **Extended Data Fig. 6 b-e**. MD-predicted de-coherence rates of NH$_4^+$ rotations are in accordance with the experimentally estimated orientational relaxation times for both the monoclinic and orthorhombic phases (see **Fig. 4f**).

We move to the collective atomic dynamics. The phonon density of state (DOS) of NH$_4$SCN is plotted from 300 up to 410 K in **Fig. 4g**. At 300 K, it is clear to see that there are three broad peaks in the profile, located at about 10, 25, and 40 meV, respectively. At 370 K, just above $T_t$, the stronger peak at 40 meV is suppressed and broadened to be featureless. At the orthorhombic-to-tetragonal phase transition, however, the tetragonal phase exhibits an almost identical profile, similar to the situation of QENS. To understand such a temperature dependence, the MD simulations are performed to reproduce the phonon DOS. Shown in **Fig. 4h** and **i** are the simulated total and partial phonon DOS for the monoclinic and orthorhombic phases, respectively. It can be seen that the low-energy phonons are attributed to the motions of SCN$^-$, while the vibrations of NH$_4^+$ mostly appear around 40 meV. The structure of partial phonon DOS is well supported by the data of the deuterated sample shown in **Extended Data Fig. 3**, where the peak is softened from 40 to about 30 meV, indicative of its NH$_4^+$ origin. As for the orthorhombic phase, the partial phonon DOS of NH$_4^+$ is much more broadened than that of SCN$^-$. Also, the peak at 40 meV is suppressed. These results are in good agreement with the experimental observations. In other words, the NH$_4^+$ vibrations are suppressed by the preferred cubic tumbling mode, which reflects the strong orientation-vibration coupling.

After the crystal structures and atomic dynamics under ambient pressure are well understood, we continue to explore the responses of the materials to external pressures, to which the inverse CBCE is directly related. Pressure-dependent synchrotron X-ray diffraction measurements indicate that the application of pressure, about 0.65 GPa, induces a phase transition from monoclinic to orthorhombic phases at room temperature. Just as shown in **Fig. 5a** and **b**, fewer Bragg peaks are observed at 0.65 GPa, indicative of higher crystal symmetry. Similarly, the pressure-dependent QENS measurements are more straightforward to testify the variation of the disorder. As shown in **Fig. 5c**, there is a much wider QENS component superposed underneath the elastic peak at about 300 MPa as compared to the ambient pressure case. In terms of the reorientation dynamics determined in the QENS analysis, the preferred cubic tumbling mode is much faster than the tetrahedral tumbling mode. Thus, the in-situ QENS measurements directly confirm the pressure-induced evolution of orientational disorder. As is known that atomic disorder is usually suppressed by pressure, the pressure-enhanced atomic disorder in $NH_4SCN$ is remarkably special.

The pressure-induced monoclinic-to-orthorhombic phase transition can be well understood by the MD-predicted atomic dynamics. The gradual application of external pressure induces localization of intense stress exclusively on $SCN^-$ ions (**Extended Data Fig. 7**). As atomic stress is a measure of instability of the system, the pressures tend to destabilize the monoclinic phase as manifested by the excited transverse vibrational mode of $SCN^-$ (see **Fig. 5d** as well as **Extended Data Fig. 8**). The mean square displacement (MSD) reflecting the vibration amplitude exhibits a much more broadened probability distribution along the direction perpendicular to $SCN^-$ under 200 MPa. The transverse vibrations of $SCN^-$ have larger components along the hydrogen-bonding directions, as manifested in the angular distribution shown in **Fig. 5e** and **f**. Pressures promote these transverse vibrations and weaken the hydrogen-bonding network of the monoclinic phase, and eventually drive the phase transition to the more stable orthorhombic phase under higher pressures through tilting of $SCN^-$ and disordering of $NH_4^+$.

Above, we have demonstrated and validated the barocaloric thermal batteries. Their realization doesn't only represent an emergent application of barocaloric materials beyond the conventional solid-state refrigeration but also brings an exciting prospect towards rational manipulation of thermal energy. As far as efficiency is concerned, the barocaloric thermal batteries are extraordinarily appealing. The input mechanical energy is estimated to be 3.85 J g$^{-1}$ by the product of the driving pressure and the induced volume changes of the materials at the phase transition, whereas stored heat is about 43 J g$^{-1}$. At full discharging, the overall efficiency is about 92% if the energy input for discharging is ignored. With the merits of high-efficiency and equilibrium-state nature, three promising applications are specified (summarized in **Supplementary Fig. 1 – Fig. 3**). First of all, it is known that thermal transfer is a key issue for solid-state refrigeration technology. With a hybrid caloric refrigeration cycle involving a regular caloric material as a refrigeration working substance and a barocaloric thermal battery as a low-temperature sink, it is possible to accelerate the solid-state thermal transfer. Secondly, in principle, pressurization needs specialized infrastructures while depressurization is much easier. Thus, barocaloric thermal batteries are suitable for circumstances of centralized charging and distributed discharging. For instance, industrial waste heat can be well collected and stored for residential heating. It is also likely to manage solar thermal energy through a diurnal cycle for building thermal management. Lastly, but most importantly, barocaloric thermal batteries provide an alternative approach to solve the intermittency problem of solar and wind electricity. In the peak of supply, electricity can be converted into heat and stored for a flexible period. In particular, barocaloric thermal batteries can be incorporated into the current compressed air energy storage system, where the thermal energy and mechanical energy are simultaneously stored. Note that, in thermodynamic analogy, such an idea of thermal batteries can be well translated to other types of caloric materials.

To summarize, the inverse CBCE has been discovered in NH$_4$SCN with the simultaneous realization of large $\Delta S_{P_0 \to P}^{max}$ and RCP. This unique property has been well understood and the fundamental microscopic mechanism has been successfully established. Rooted in this effect, the high-performance barocaloric thermal batteries are materialized, which are characteristic of high thermodynamic stability, fast pressure response, and excellent tunability. The demonstrated barocaloric thermal batteries are expected to benefit rational energy utilization.

**Acknowledgments**

The work conducted in the Institute of Metal Research was supported by the Key Research Program of Frontier Sciences of Chinese Academy of Sciences (Grant no. ZDBS-LY-JSC002), the Ministry of Science and Technology of China (Grant no. 2020YFA0406002), the Liaoning Revitalization Talents Program (Grant no. XLYC1807122) and National Natural Science Foundation of China (Grant no. 11804346). S.L. acknowledges the startup grant and financial support from the Oceanic Interdisciplinary Program (Grant no. SL2020MS008) from Shanghai Jiao Tong University. K.L. acknowledges the support by the National Natural Science Foundation of China (Grant nos. 21771011 and 22022101). B.Y. acknowledges the Ministry of Science and Technology of China (Grant no. 2016YFA0401503). We acknowledge the beam time provided by SPring-8 (Proposal no. 2019A2052), CSNS (Proposal no. P1819062700003), ANSTO (Proposal no. 7867), and BSRF. We also acknowledge the excellent support from Amy Shumack and Rachel White of the Sample Environment Group at ANSTO for getting approval on using high-pressure cell and subsequent setting up the high-pressure equipment on the Pelican instrument.

**Author Contributions**

B.L. proposed the project. Zhe Zhang was co-supervised by B.L. and Zhidong Zhang. K.L. and Y.W. prepared and characterized the deuterated samples. Zhe Zhang and B.L. collected heat flow data of all samples. Zhe Zhang, R.S., K.L. and Y.W. measured the pressure-induced temperature changes. Zhe

Zhang, J.C., and L.H. collected the neutron powder diffraction data. S.K. conducted synchrotron X-ray diffraction under ambient pressure. K.L. and X.L. performed pressure-dependent synchrotron X-ray diffraction. Zhe Zhang and K.L. refined all diffraction data. D.Y., Zhe Zhang, Zhao Zhang, C.Y., and B.L. performed INS and QENS measurements. D.Y and R. M commissioned the high-pressure cell on the Pelican instrument. The high-pressure cells were provided by B.Y., Q.R., and X.T. Zhe Zhang and B.L. analyzed the INS and QENS data. S.L., J.W., K.Q., Z.C., J.Y., and M.W. conducted DFT and MD simulations for lattice parameters, reorientation dynamics, hydrogen bonding, phonon density of states, and local atomic stress analyses. C.Z. estimated the efficiency of the barocaloric thermal batteries. R.S. and B.L. conducted finite-element thermal simulations. B.L. and Zhe Zhang wrote the manuscript with inputs from all authors.

**Data availability**

The data of the present study are available from the corresponding authors upon reasonable request.

**Competing interests**

The authors declare no competing interests.

**Additional information**

Supplementary information is available for this paper at XXX.

**Methods**

**Samples**

NH4SCN was directly purchased from Aladdin. ND4SCN was prepared by replacing H with D in an aqueous solution of NH4SCN in $D_2O$. 5 grams of NH4SCN were dissolved completely in 30 mL of $D_2O$. After keeping static for 20 minutes, $H_2O$ and redundant $D_2O$ were removed using a rotary evaporator. The product was dried at 333 K for 6 hours under a vacuum. Such procedures were repeated several times to improve the purity. The quality of the products was confirmed by X-ray diffraction and Raman scattering (**Extended Data Fig. 3**). All the samples were kept in a glove box and dehydrated for 1 hour at a temperature of 348 K before all the relevant experimental measurements.

**Pressure-induced temperature changes**

The pressure-induced temperature changes based on room temperature were measured on a VX3 Paris-Edinburgh press device with a K-typed thermocouple. NH4SCN powder samples weighted about 120 mg were pre-pressed into two hemispheres, where the thermocouple was sandwiched, and then they were placed onto the double-toroidal tungsten carbide (WC) anvils. An automatic hydraulic oil syringe pump was used to control the pressurization and depressurization processes at 5, 10, and 20 ml min$^{-1}$. The maximum pressure was up to 300 MPa. NaCl was also measured as a reference. For more details, please refer to **Supplementary Fig. 4 – Fig. 6** as well as **Supplementary Video 2**.

**Pressure-dependent calorimetric characterizations**

The heat flow data were collected as a function of temperature and pressure using a high-pressure differential scanning calorimeter (μDSC7, Setaram). The sample was enclosed in a high-pressure vessel made of Hastelloy. Hydrostatic pressure up to 100 MPa was achieved by compressing nitrogen gas and accurately controlled by a high-pressure panel. Constant pressure scans were performed on NH4SCN at 0.1, 20, 40, 60, 80, and 100 MPa at a rate of 1 K min$^{-1}$, respectively. Heat flow data after subtracting baseline background were converted to entropy changes as described before[9]. RCP was obtained as follows[22],

$$\text{RCP} = \Delta S_{P_0 \to P}^{max} \cdot \delta T_{\text{FWHM}} \tag{1}$$

where $\delta T_{\text{FWHM}}$ is the full width at half maximum for $\Delta S_{P_0 \to P}^{max}$ as a function of temperature. For the constant temperature process, the samples of $NH_4SCN$ were pressurized from 8.2 to 90 MPa at 345, 349, and 355 K and depressurized back to 8.2 MPa at 318, 313, and 315 K, respectively. Under the same conditions, the signals of an empty cell were obtained and subtracted as background.

**Neutron powder diffraction**

Neutron powder diffraction experiments were performed on $ND_4SCN$ at the general purpose powder diffractometer (GPPD)[31] of China Spallation Neutron Source (CSNS) in China. The powder amount of 0.57 gram was put into a vanadium can ($\phi$ = 9 mm, $L$ = 70 mm). Constant-temperature scans were done at 300 K, 375, and 400 K, respectively. At each temperature, the counting persisted for 5 hours. The diffraction patterns were fitted based on individual structural models (listed in **Extended Data Table 1**) using Jana2006 software[32].

**Synchrotron X-ray diffraction**

Ambient pressure synchrotron X-ray diffraction was conducted at the beamline BL02B2 of SPring-8 in Japan[33]. The X-ray wavelength was 0.999108 Å. A small amount of powder sample was sealed into a quartz capillary ($\phi$ = 0.5 mm). The diffraction data were obtained in a temperate region from room temperature up to 380 K. Each scan took about 3 minutes. The pressure-dependent diffraction data were obtained at the high-pressure beamline of the Beijing Synchrotron Radiation Facility (BSRF) in China. The standard sample of $CeO_2$ was used for calibration before the experiment. The powder sample was loaded into a diamond anvil cell to access pressure of 0.07 and 0.65 GPa at room temperature. The resulting data were analyzed by Le Bail fitting in Jana2006 software[32].

**INS and QENS**

The INS and QENS experiments of NH4SCN and ND4SCN were conducted at the time-of-flight neutron spectrometer Pelican of Australian Center for Neutron Scattering (ACNS) of Australian Nuclear Science and Technology Organisation (ANSTO) in Australia[34]. The instrument was configured for incident neutron wavelengths of 2.345, 4.69, and 5.96 Å, corresponding to incident energy of 14.9, 3.7, and 2.3 meV with resolutions of 0.7, 0.135, and 0.065 meV at the elastic lines, respectively. The dehydrated samples were sealed into aluminum cans under a helium gas environment. The experiments were performed in a temperature region from room temperature up to 410 K for NH4SCN and at 300 and 380 K for ND4SCN. The pressure-dependent measurements were carried out at 0.1 and 300 MPa with an incident wavelength of 4.69 Å at 300 K using a high-pressure cell made of high-strength aluminum alloy. The background was corrected by the data of empty cans measured under the same conditions. The instrument resolution functions were obtained on a standard cylindrical vanadium tube at 300 K. The spectra of the vanadium standard were also used for detector normalization. The data reduction and processing were completed in the Large Array Manipulation Program (LAMP) software[35]. The scattering function $S(Q,E)$, as a function of scattering wave vectors ($Q$) and energy transfer ($E$), were measured over a wide temperature range. The generalized PDOS were obtained using the formula,

$$g(E) = \int \frac{E}{Q^2} S(Q,E) \left(1 - e^{-\frac{E}{k_B T}}\right) dQ. \tag{2}$$

where $k_B$ is Boltzmann's constant and $T$ is temperature. The $S(Q,E)$ data were sliced at selected $Q$-points, where Bragg peaks were avoided. The sliced spectra were fitted in the PAN module of Data Analysis and Visualization Environment (DAVE)[36]. One Lorentzian function, one delta function, and a linear background were used for the spectral fitting with wavelengths of 4.69 Å and 5.96 Å. One more damped harmonic oscillator function was added to fit the spectra with a wavelength of 2.345 Å because phonons were probed. EISF was defined by the intensity ratio[37],

$$\text{EISF} = \frac{I_{\text{elastic}}}{I_{\text{elastic}} + I_{\text{QENS}}} \tag{3}$$

Four types of NH4$^+$ reorientation models were considered, as follows,

$C_2$ or $C_3$ uniaxial jumps[37],

$$\text{EISF} = \frac{1}{2}[1 + j_0(Qd_{\text{H-H}})] \tag{4}$$

Tetrahedral tumbling[38],

$$\text{EISF} = \frac{1}{4}[1 + 3j_0(Qd_{\text{H-H}})] \tag{5}$$

Cubic tumbling[39],

$$\text{EISF} = \frac{1}{8}[1 + 3j_0\left(\frac{2}{\sqrt{3}}Qd_{\text{N-H}}\right) + 3j_0(Qd_{\text{H-H}}) + j_0(2Qd_{\text{N-H}})] \tag{6}$$

Isotropic rotational diffusion[37],

$$\text{EISF} = j_0^2(Qd_{\text{N-H}}) \tag{7}$$

Where N–H bond distance $d_{\text{N–H}} = 1.02$ Å and H–H bond distance $d_{\text{H–H}} = (8/3)^{1/2} d_{\text{N–H}}$, and $j_0(x) = \frac{\sin(x)}{x}$[28,37], which is the spherical Bessel function of the zeroth-order.

**DFT and MD simulations**

We conducted density functional theory (DFT) calculations using VASP[40,41] to determine the lattice parameters of NH$_4$SCN at orthorhombic and monoclinic phases. The crystal structures obtained from neutron diffraction experiments on deuterated samples were used as the initial structure. The projector augmented wave (PAW)[42] pseudopotentials are utilized to describe core and valence electrons, and the generalized gradient approximation based on the Perdew-Burke-Ernzerhof (GGA-PBE)[43] function is selected to describe electron exchange and correlation. Since the standard GGA approximation underestimates van der Waals interactions, the latest Grimme's energy correction method is employed to correctly account for these dispersion interactions[44]. We select the plane-wave based kinetic energy cutoff of 550 eV, and the $\Gamma$-centered 6×6×4 Monkhorst-Pack[45] $k$-point mesh for sampling the Brillouin zones of the crystal, as to minimize atomic forces below 1.0 meV Å$^{-1}$. Lattice parameters were then obtained from DFT-optimized structures (**Extended Data Table 2**).

All-atomistic MD simulations were carried out using LAMMPS[46] to determine the lattice parameters of monoclinic and orthorhombic phases (**Extended Data Table 2**). The GAFF-ESP (General AMBER Force Field with charges derived using the electrostatic-potential method) force field[47,48] parameters defined in **Supplementary Table S1 and S2** were employed in our study. The time step was set to 1 fs for all MD simulations. A cutoff distance of 1 nm for the short-range Lennard-Jones (LJ) and Coulombic interactions was chosen. The particle mesh Ewald (PME) summation method[49] was used here to treat long-range electrostatic interactions beyond the cutoff. Three-dimensional periodic boundary conditions were applied to all the simulations. All MD equilibrations of the initial structures were carried out for 2 ns under the *NPT* ensemble after energy minimization.

To provide transient information on reorientational dynamics of $NH_4^+$, we calculated the averaged cosine trajectory of N-H bonds using 40 ps of MD simulations after the equilibration procedure above. Using atomic trajectories saved every 200 fs, we calculated the average cosine trajectory $\cos\theta_i(t-t_0)$ over all the $NH_4^+$ cations, *i.e.*, the angle spanned by the N-H bond vector as a function of time, $t$:

$$\cos\theta_i(t-t_0) = \hat{n}_i(t)\cdot\hat{n}_i(t_0) \tag{8}$$

where $\hat{n}_i(t)$ is the unit vector for the N-H bond at any time $t$ and $\hat{n}_i(t_0)$ is that at the initial time $t_0 = 20$ ns. Based on the time needed for $\cos\theta_i(t-t_0)$ to decay to zero, we estimated the de-coherent lifetime for $NH_4^+$ rotations (see **Extended Data Fig. 6**).

The phonon DOS was computed for the orthorhombic and monoclinic phases of $NH_4SCN$ at their corresponding temperatures (350 and 300 K) and pressures (0.1 and 200 MPa). After the equilibration procedure mentioned above, a 10 ps simulation was performed for each model, during which velocities of the atoms were collected every 2 fs to calculate the velocity autocorrelation function (VACF). The phonon DOS was calculated by taking the discrete Fourier transform of the mass-weighted VACF. Specifically[50],

$$D(\omega) = \int_0^\tau \left(\sum_{i=1}^N m_i \frac{\langle v_i(t)\cdot v_i(0)\rangle}{\langle v_i(0)\cdot v_i(0)\rangle}\right) e^{-i\omega t} dt \tag{9}$$

where $\omega$ is the phonon angular frequency; $m_i$ is the mass of atom $i$; $v_i(t)$ and $v_i(0)$, respectively, are the velocities of atom $i$ at time $t$ and 0; and $N$ is the total number of atoms. Contributions to the total phonon DOS from $NH_4^+$ and $SCN^-$ were calculated similarly. To characterize the pressure-induced changes in the transverse and longitudinal vibrational modes of $SCN^-$, the MSD functions of $SCN^-$ along the transverse (radial) and longitudinal (axial) directions of $SCN^-$ vector in a cylindrical coordinate were collected every 10 fs during a 1 ns MD simulation.

Using MD-predicted atomic trajectories, atomic von Mises stresses are computed to quantify the local stress to cover both normal and shear components. Based on the symmetric Cauchy stress tensor theory[51], the second invariant of the stress tensor is defined as the von Mises stress[52,53]:

$$\sigma_{vm} = \sqrt{\frac{(\sigma_{xx} - \sigma_{yy})^2 + (\sigma_{yy} - \sigma_{zz})^2 + (\sigma_{xx} - \sigma_{zz})^2 + 6(\sigma_{xy}^2 + \sigma_{yz}^2 + \sigma_{zx}^2)}{2}} \quad (10)$$

where $\sigma_{xx}$, $\sigma_{yy}$ and $\sigma_{zz}$ are the per-atom normal stresses along $x$, $y$, and $z$-axis, respectively, defined by the virial theory. Similarly, $\sigma_{xy}$, $\sigma_{yz}$, and $\sigma_{zx}$ are the per-atom shear stresses. The OVITO software package[54] is used to display local stress information.

**FE thermal simulations**

The simulations were implemented based on the solid mechanics and heat transfer modules of COMSOL Multiphysics platform[55]. The parameters of physical properties of NH4SCN were listed in **Supplementary Table 3**. A two-dimensional axisymmetric finite element model with a press die was used to simulate the heat flow data, whose parameters were summarized in **Supplementary Table S4** and **S5**. Following the experimental setup, a semicircle model was used to simulate the pressure-induced temperature changes, whose diameter is 6 mm. The phase transition process was parameterized by a step function of the phase fraction of low-temperature monoclinic phase (*f*) as a function of pressure and temperature, as follows,

$$Step(P,T) + f = 1 \quad (11)$$

$$Step(P,T) = \frac{1}{1+e^{-\frac{a}{T_1-T_2} \times [(T_1-T_2) \times P - (P_1-P_2) \times T - T_1 P_2 + T_2 P_1]}} \quad (12)$$

where a = 0.5, $P_1$ = 20 MPa, $P_2$ = 0.1 MPa, $T_1$ = 358 K, and $T_2$ = 363.83 K. The pressure-time profiles for these simulations are summarized in **Supplementary Fig. 6**. The heat transfer coefficient $h$ was set to be $5 \times 10^4$ W m$^{-2}$ K$^{-1}$ for $\Delta T$ and 100 W m$^{-2}$ K$^{-1}$ for heat flow data for the thermal interface between the barocaloric material and the environments.

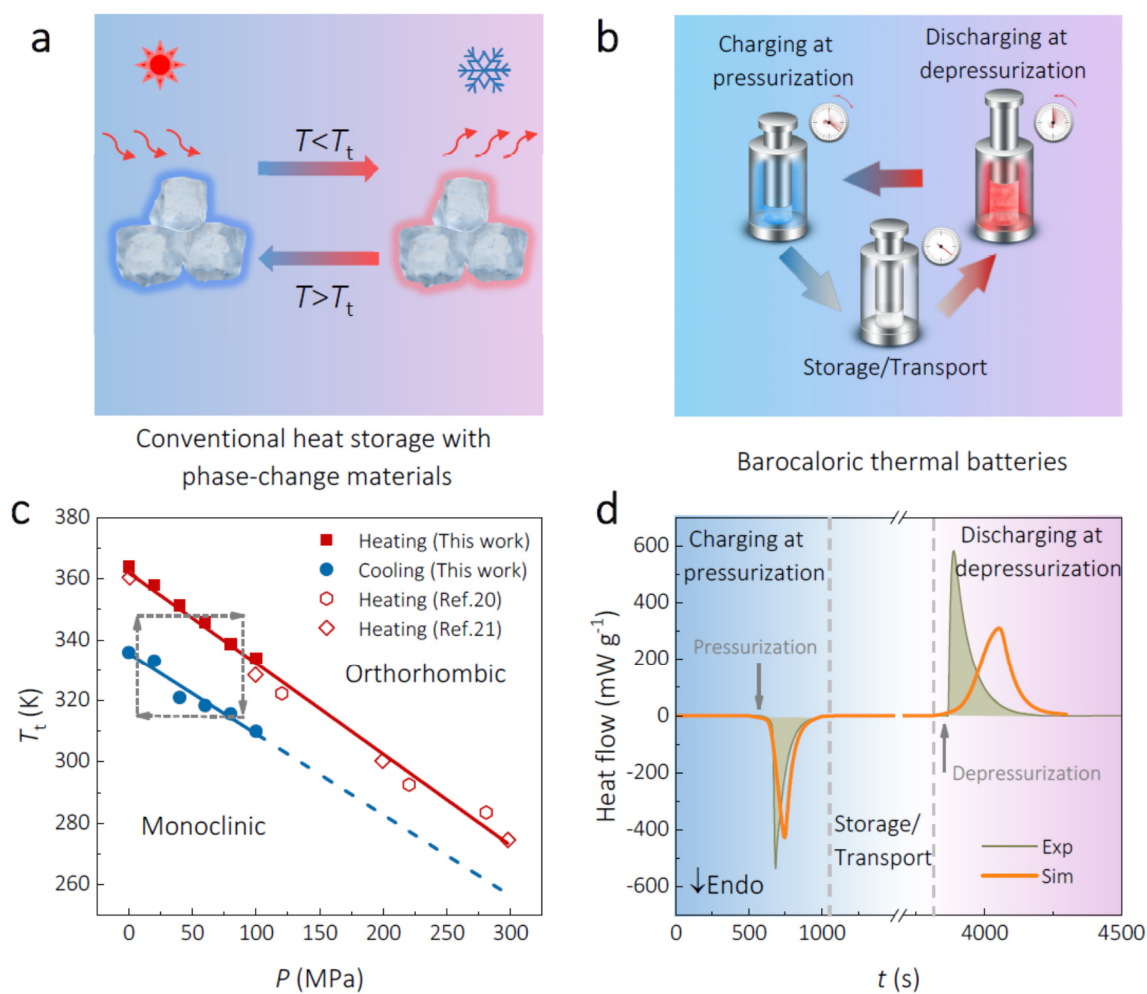

**Fig. 1 Barocaloric thermal batteries: concept and realization. a**. A schematic diagram of conventional heat storage with phase-change materials. The symbols of sun and snow denote the high-temperature and low-temperature environments. The curved arrows represent the heat transfer directions. **b**. A schematic diagram for barocaloric thermal batteries including thermal charging at pressurization, storage/transport, and thermal discharging at depressurization. **c.** Phase diagram defined by the relationship between the transition temperature ($T_t$) and applied pressures. The data points at higher pressures are obtained from refs. 20 and 21. The dash lines describe a complete process for barocaloric thermal batteries. **d**. Heat flow variations as a function of time for a barocaloric thermal battery process (dash lines in **c**, with the heating process omitted) obtained on $NH_4SCN$ in the experiment at pressures changing between 8.2 and 90 MPa and in the simulation at pressure changing between 0.1 to 200 MPa with heat transfer coefficient (*h*) of 100 W m$^{-2}$ K$^{-1}$. The moments for applying and releasing pressures are arrowed.

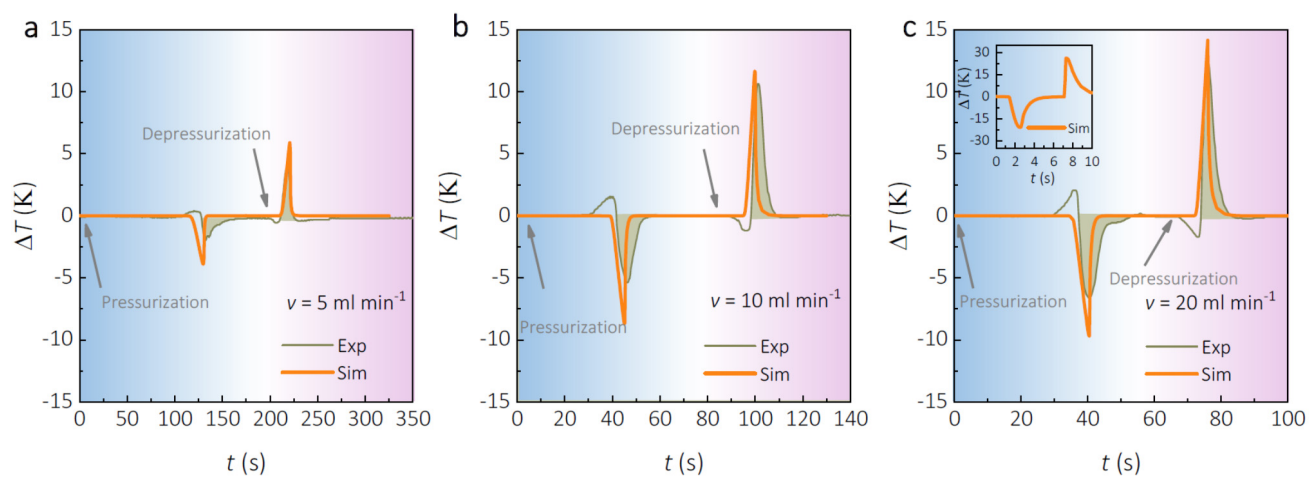

**Fig. 2 Practical performances of barocaloric thermal batteries operated at room temperature**. **a**. The experimental and simulated relative temperature change (Δ*T*) during charging (0.1 to 300 MPa) and discharging (300 to 0.1 MPa) at a ramping rate of 5 ml min$^{-1}$ (see **Methods** for details). **b**. Δ*T* at a ramping rate of 10 ml min$^{-1}$. **c**. Δ*T* at a ramping rate of 20 ml min$^{-1}$. The inset shows the simulated process at extremely fast pressure ramping, leading to a maximum Δ*T* of about 26 K, whose process is recorded in **Supplementary Video 1**.

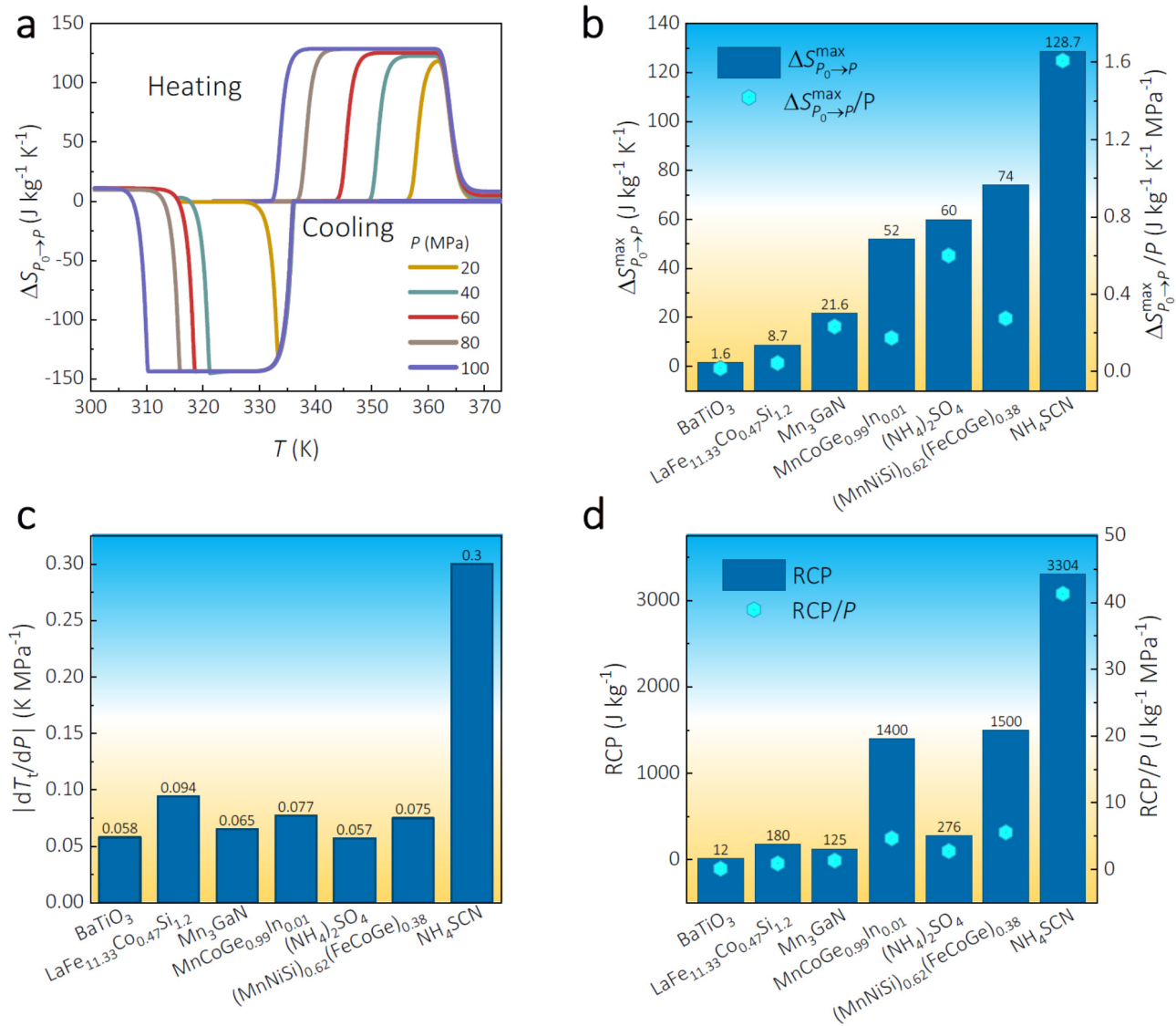

**Fig. 3 Inverse CBCE of NH4SCN. a**. Pressure-induced entropy changes ($\Delta S_{P_0 \to P}$) at pressure changing from $P_0 = 0.1$ MPa to $P = 20, 40, 60, 80,$ and $100$ MPa. **b**. Maxima of pressure-induced entropy changes ($\Delta S_{P_0 \to P}^{max}$) and normalized values by the driven pressure of NH4SCN compared with other leading materials with inverse barocaloric effects. **c**. $|dT_t/dP|$ for these materials. **d**. RCP and normalized RCP for these compounds.

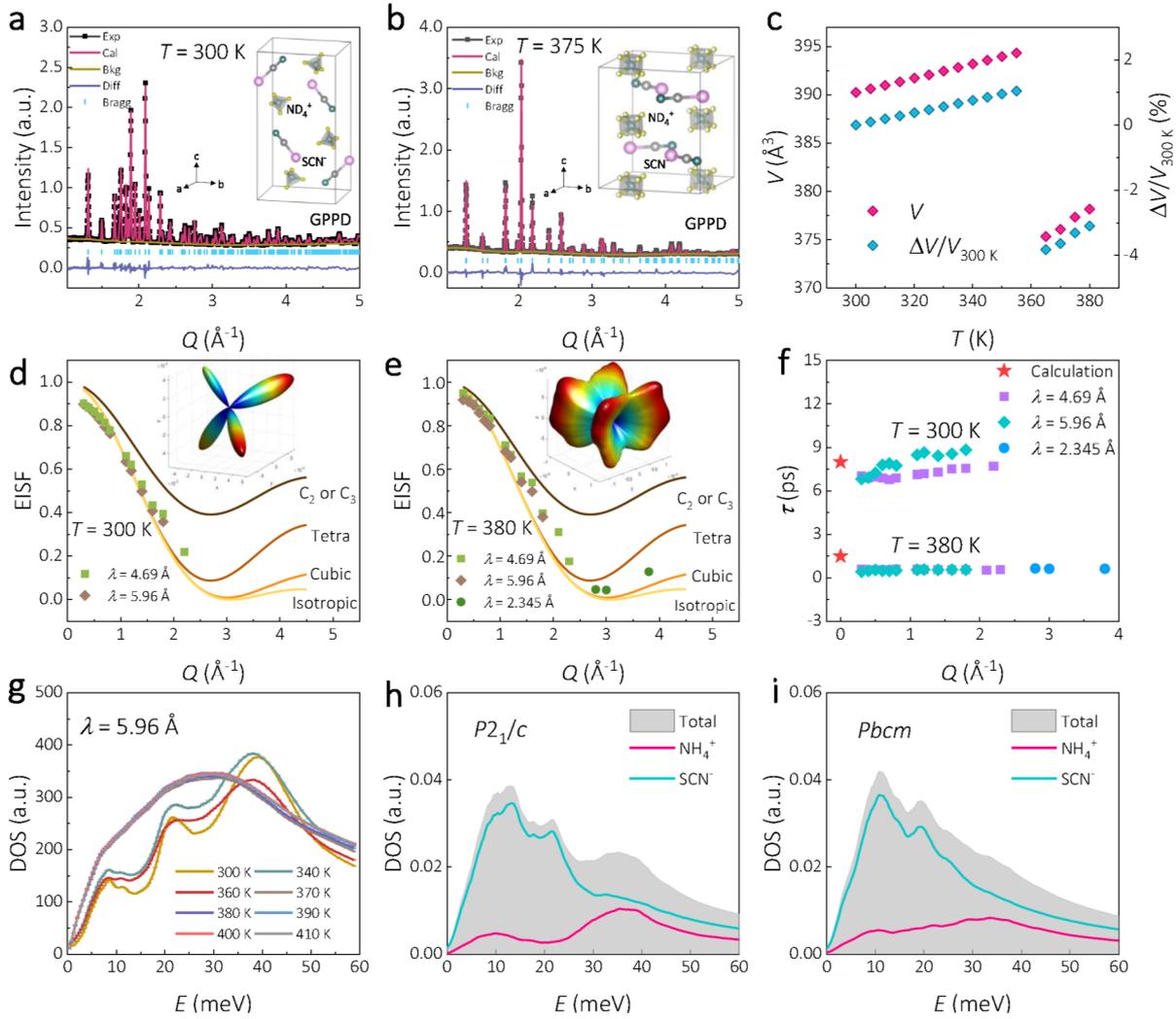

**Fig. 4 Crystal structures and atomic dynamics under ambient pressure. a** and **b**. Rietveld refinements of the neutron powder diffraction patterns of ND$_4$SCN at 300 K based on the $P2_1/c$ model and at 375 K based on the *Pbcm* model (in the insets). For the fitting, $R_{wp}$ = 3.41%, $R_p$ = 2.77% and GOF = 4.57; $R_{wp}$ = 3.89%, $R_p$ = 3.02% and GOF = 5.16, respectively. The detailed structural parameters determined in the refinements are listed in **Extended Data Table 1**. **c**. Unit cell volume and their relative changes as a function of temperature determined based on synchrotron X-ray diffraction data (**Extended Data Fig. 4**). **d** and **e**, Experimental EISF compared to a few models at 300 and 380 K, respectively. The insets are the spatial distribution of hydrogen atoms of NH$_4^+$ determined in MD simulations. **f**. Experimental relaxation time ($\tau$) compared to the MD simulations (**Extended Data Fig. 6**). **g**. Experimental DOS up to 60 meV at different temperatures. **h** and **i**, MD-predicted phonon DOS for $P2_1/c$ and *Pbcm* phases, respectively.

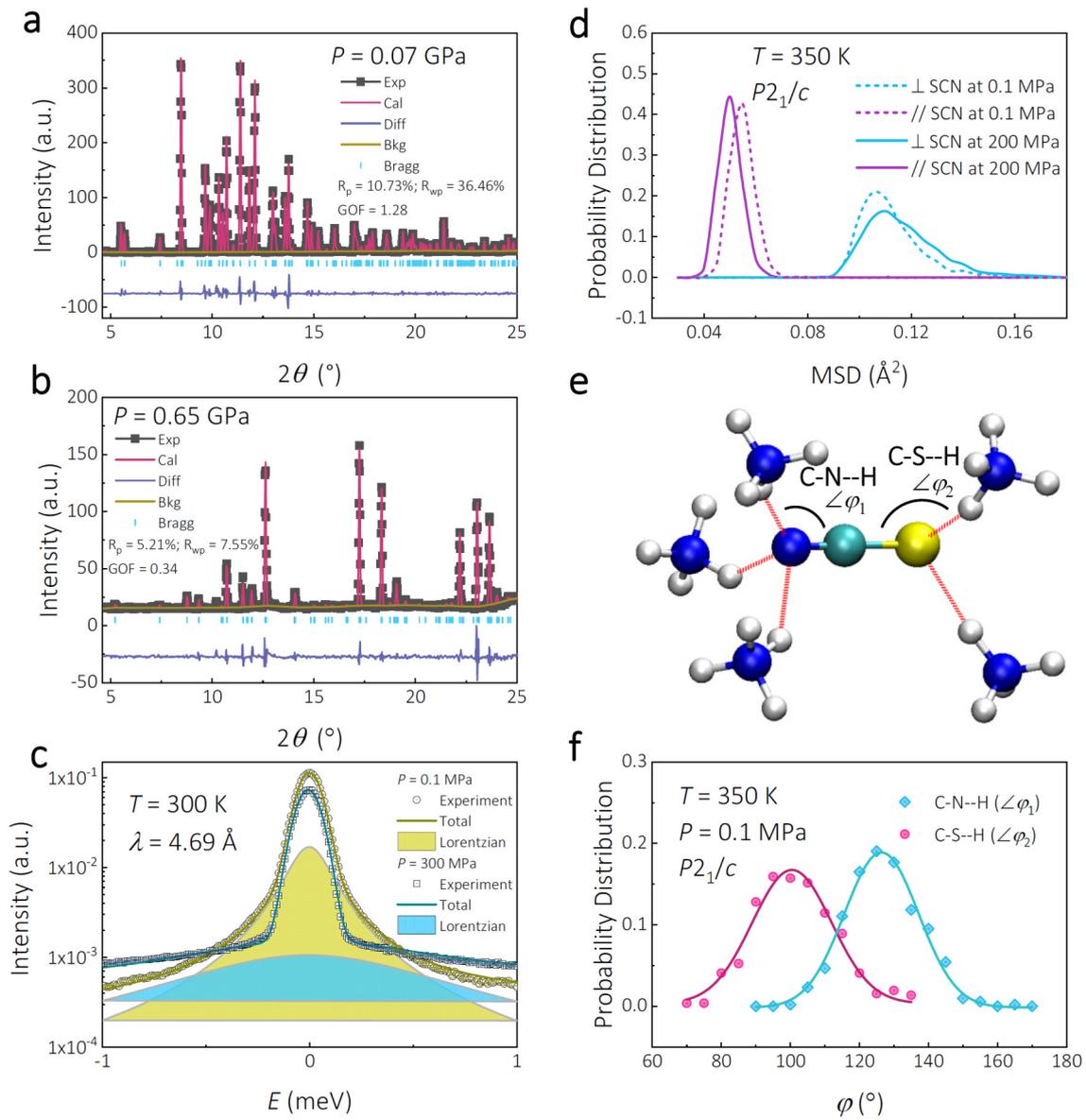

**Fig. 5 Structural and dynamics in response to external pressures. a** and **b**, Synchrotron X-ray diffraction patterns and refinements at room temperature under 0.07 and 0.65 GPa, respectively. **c**. $Q$-integrated QENS spectra at room temperature under ambient pressure and 300 MPa. The Lorentzian components are highlighted as shaded regions. **d**. Probability distribution of MSD (amplitude of atomic vibrations) decomposed into the directions perpendicular to and parallel with SCN⁻ vectors in the $P2_1/c$ model under ambient pressure and 200 MPa at 350 K. **e**. Illustration of C-N--H and C-S--H hydrogen bonds, whose angles with respect to the SCN⁻ vector are labeled as $\varphi_1$ and $\varphi_2$, respectively. **f**. Probability distribution of angles $\varphi_1$ and $\varphi_2$. The solid lines represent fits to Gaussian functions, giving rise to peaks at 100.5(6) and 126.4(3)°.

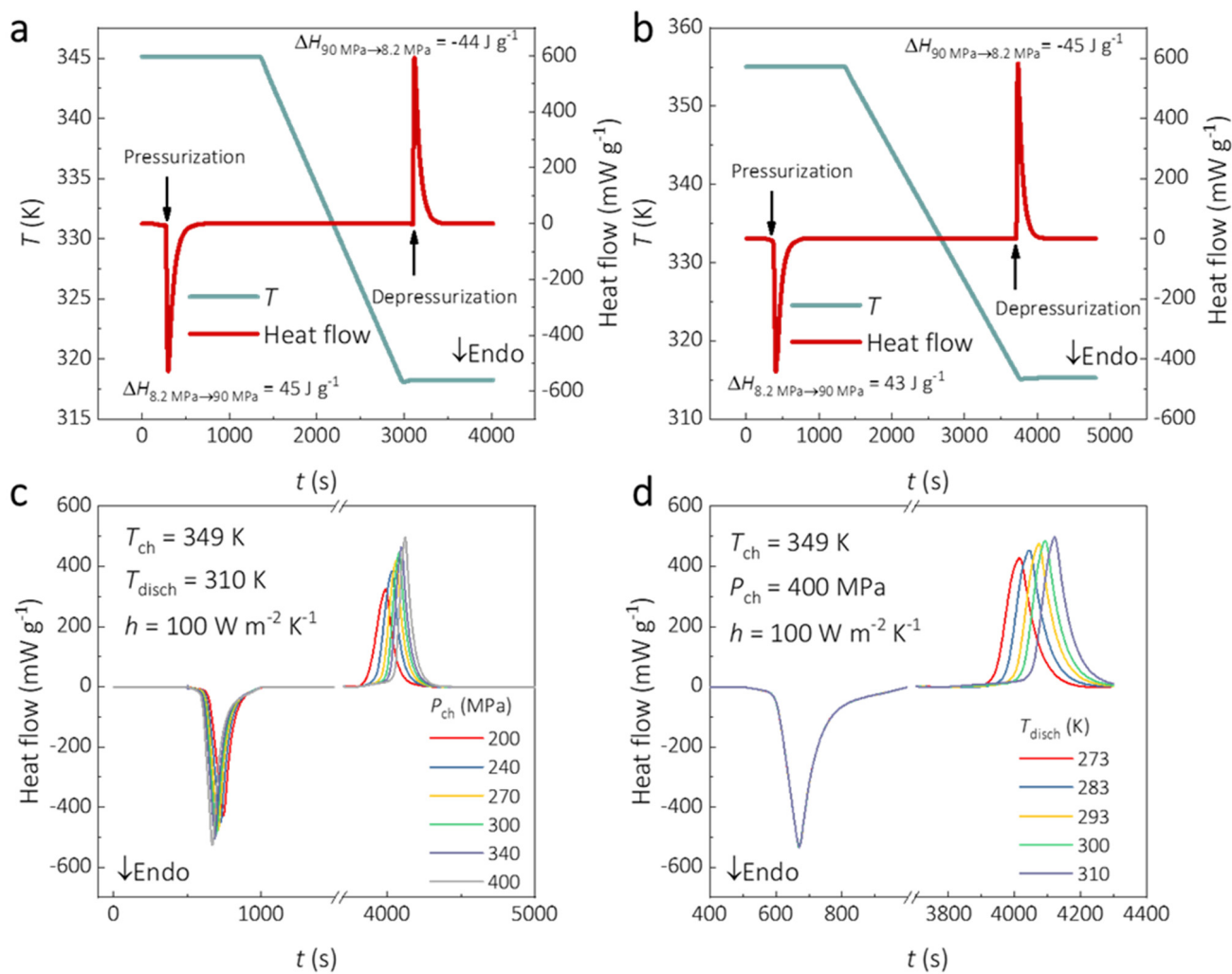

**Extended Data Fig. 1. Additional charging and discharging data**. **a** and **b**, Heat flow variation in two charging-discharging processes obtained in NH$_4$SCN. The moments for applying and releasing pressures are arrowed. The charging temperatures are 345 and 355 K while the discharging temperatures are 318 and 315 K for **a** and **b**, respectively. **c**. Simulated charging-discharging process at charging temperature of 349 K, charging pressures of 200, 240, 270, 300, 340, and 400 MPa, discharging temperatures of 310 K, and thermal transfer coefficient ($h$) of 100 W m$^{-2}$ K$^{-1}$. **d**. Simulated charging-discharging process at charging temperature of 349 K, charging pressure of 400 MPa, discharging temperatures of 310, 300, 293, 283 and 273 K, and $h$ of 100 W m$^{-2}$ K$^{-1}$.

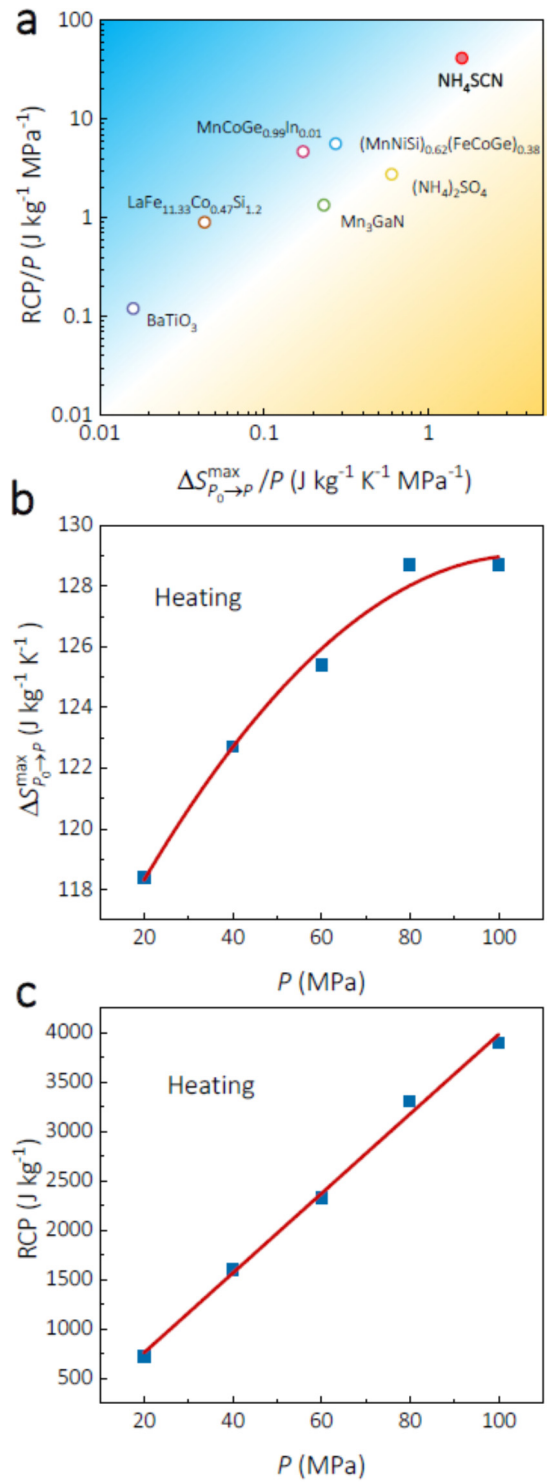

**Extended Data Fig. 2. Additional data of inverse CBCEs**. **a**. Normalized RCP and maximum entropy changes of these materials. NH4SCN is the best. **b** and **c**, Pressure dependence of maximum entropy changes and RCP of NH4SCN.

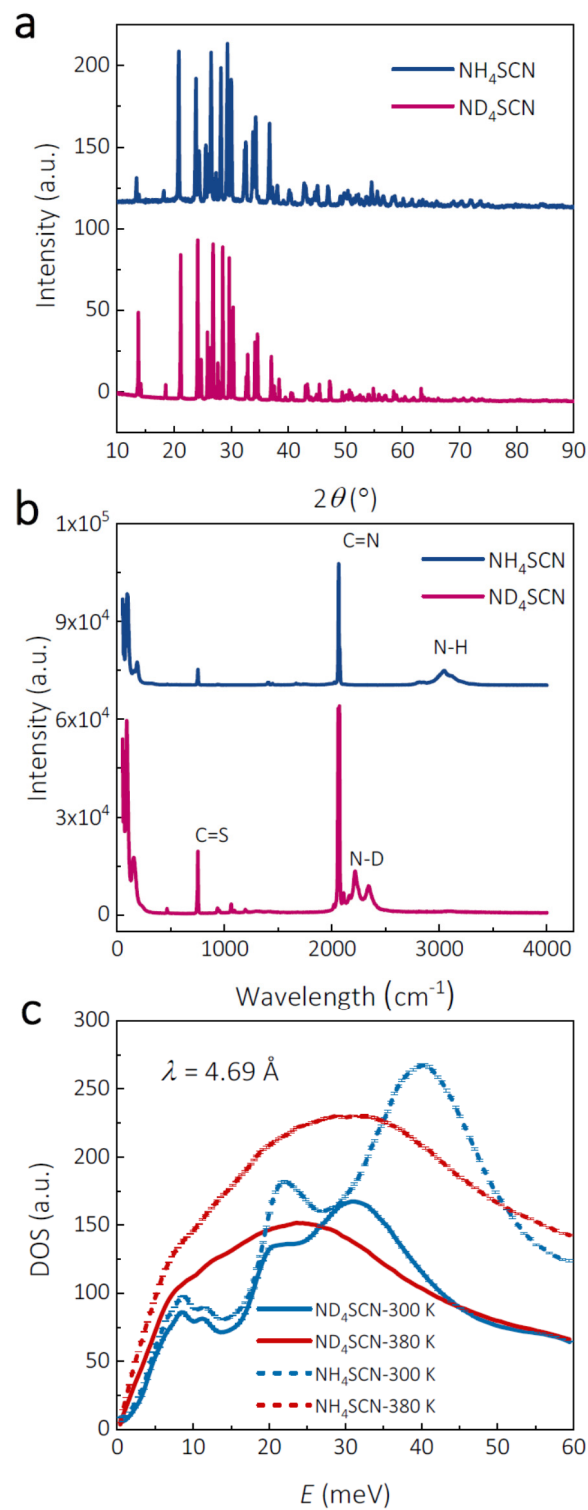

**Extended Data Fig. 3. Characterization of deuterated samples. a**. Comparison of laboratory XRD data of ND₄SCN with NH₄SCN. **b**. Raman scattering data of NH₄SCN and ND₄SCN at room temperature. **c**. DOS of ND₄SCN and NH₄SCN at 300 K and 380 K with the incident neutron wavelength of 4.69 Å.

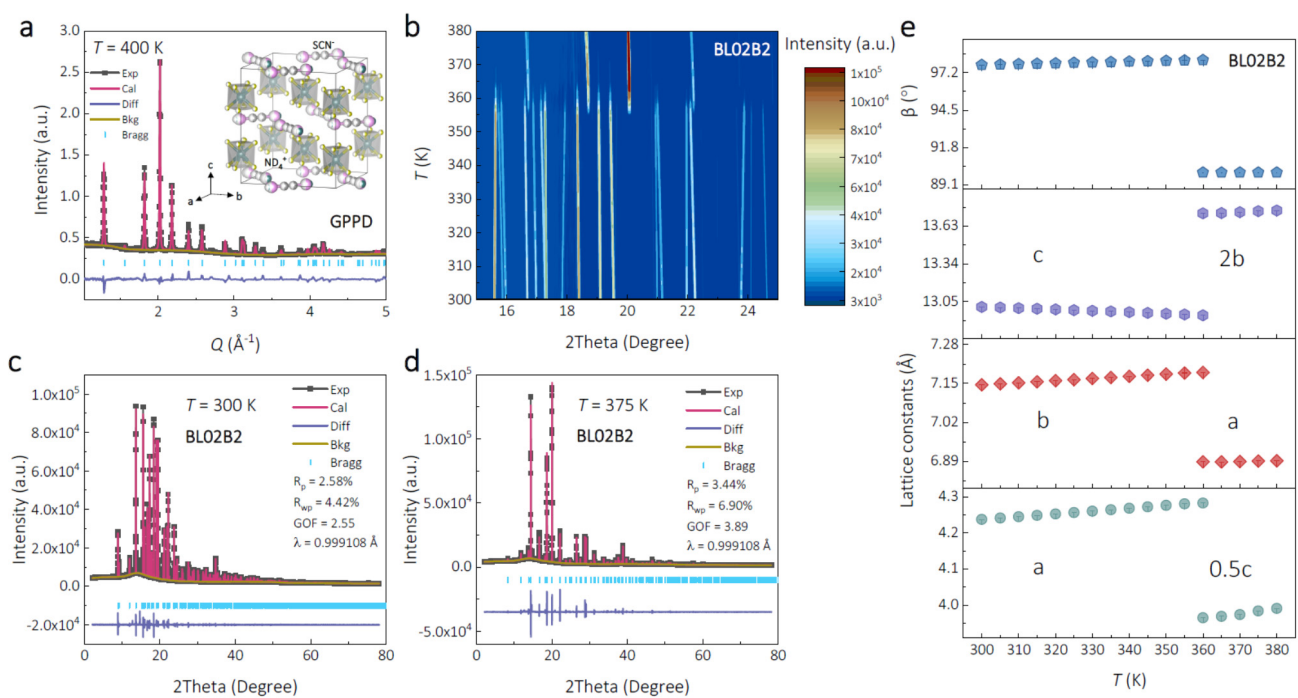

**Extended Data Fig. 4. Additional neutron and synchrotron X-ray diffraction data. a**. Rietveld refinement on the neutron powder diffraction pattern of ND$_4$SCN at 400 K based on the *I*4/*mcm* model shown in the inset. $R_{wp}$ = 3.45%, $R_p$ = 2.22%, GOF = 4.57. **b**. Ambient pressure synchrotron X-ray diffraction of NH$_4$SCN under varying temperatures. **c** and **d**, Synchrotron X-ray diffraction patterns and refinements at the temperatures of 300 and 375 K. **e**. Temperature dependence of unit-cell parameters determined based on synchrotron X-ray diffraction data.

**Extended Data Table 1. Experimental structural parameters of three phases determined in neutron diffraction on ND$_4$SCN at GPPD.**

| Atoms | x | y | z | Occ | $U_{iso}$ |
|---|---|---|---|---|---|
| *T* = 300 K, monoclinic, space group: *P*2$_1$/*c* | | | | | |
| *a* = 4.22430(16) Å, *b* = 7.1294(3) Å, *c* = 12.9864(5) Å, *β* = 97.751(3)° | | | | | |
| S  | 0.029(4)    | 0.989(2)    | 0.2017(13) | 1 | 0.004(6) |
| C  | 0.1116(18)  | 0.8167(13)  | 0.1151(7)  | 1 | 0.016(3) |
| N1 | 0.1632(15)  | 0.7069(9)   | 0.0572(4)  | 1 | 0.030(2) |
| N2 | 0.4420(8)   | 0.3353(5)   | 0.1092(3)  | 1 | 0.030(2) |
| D1 | 0.5677(19)  | 0.2925(12)  | 0.0517(4)  | 1 | 0.095(4) |
| D2 | 0.5962(18)  | 0.3628(12)  | 0.1749(4)  | 1 | 0.086(5) |
| D3 | 0.2873(16)  | 0.2320(10)  | 0.1241(6)  | 1 | 0.081(4) |
| D4 | 0.317(2)    | 0.4537(9)   | 0.0861(8)  | 1 | 0.121(6) |
| *T* = 375 K, orthorhombic, Space group: *Pbcm* | | | | | |
| *a* = 6.867(3) Å, *b* = 6.853(3) Å, *c* = 7.934(3) Å | | | | | |
| S  | 0.357(4)    | 0.094(3)    | 0.25       | 1   | 0.060(3)  |
| C  | 0.2148(15)  | 0.2941(12)  | 0.25       | 1   | 0.060(3)  |
| N1 | 0.1195(15)  | 0.4289(13)  | 0.25       | 1   | 0.060(3)  |
| N2 | 0.7841(12)  | 0.25        | 0          | 1   | 0.071(3)  |
| D1 | 0.6979(13)  | 0.338(7)    | 0.074(6)   | 0.5 | 0.315(13) |
| D2 | 0.8698(13)  | 0.156(5)    | -0.066(6)  | 0.5 | 0.315(13) |
| D3 | 0.8699(13)  | 0.165(7)    | 0.075(6)   | 0.5 | 0.315(13) |
| D4 | 0.6983(13)  | 0.327(7)    | -0.081(5)  | 0.5 | 0.315(13) |
| *T* = 400 K, Tetragonal, Space group: *I*4/*mcm* | | | | | |
| *a* = 6.8810(2) Å, *b* = 6.8810(2) Å, *c* = 8.0252(4) Å | | | | | |
| S  | 0.6360(11) | 0.1360(11)  | 0         | 0.5 | 0.087(4)  |
| C  | 0.4635(9)  | -0.0365(9)  | 0         | 0.5 | 0.087(4)  |
| N1 | 0.3414(8)  | -0.1586(8)  | 0         | 0.5 | 0.087(4)  |
| N2 | 0          | 0           | 0.25      | 1   | 0.063(3)  |
| D  | 0.136(2)   | -0.017(8)   | 0.1485(15)| 0.5 | 0.453(17) |

**Extended Data Table 2. Comparison of experimental and simulated lattice parameters.**

| Lattice Parameters | Methods | $a$ (Å) | $b$ (Å) | $c$ (Å) | $\beta$ (degree) |
|---|---|---|---|---|---|
| Monoclinic | DFT | 4.10 | 7.03 | 13.17 | 96.89 |
| | MD | 3.96 | 6.27 | 13.70 | 97.77 |
| | NPD (300 K) | 4.22430(16) | 7.1294(3) | 12.9864(5) | 97.751(3) |
| | SXRD (300 K) | 4.237355(13) | 7.14534(2) | 13.00870(4) | 97.7735(2) |
| Orthorhombic | DFT | 6.75 | 6.99 | 7.58 | 88.29 Note1 |
| | MD | 6.61 | 6.60 | 7.45 | 90.00 |
| | NPD (375 K) | 6.867(3) | 6.853(3) | 7.934(3) | 90.00 |
| | SXRD (375 K) | 6.89071(5) | 6.87274(5) | 7.96787(7) | 90.00 |

Note 1: The DFT predicted $\beta$ deviates slightly from 90 degrees due to the frozen $NH_4^+$ dynamics during DFT energy minimization.

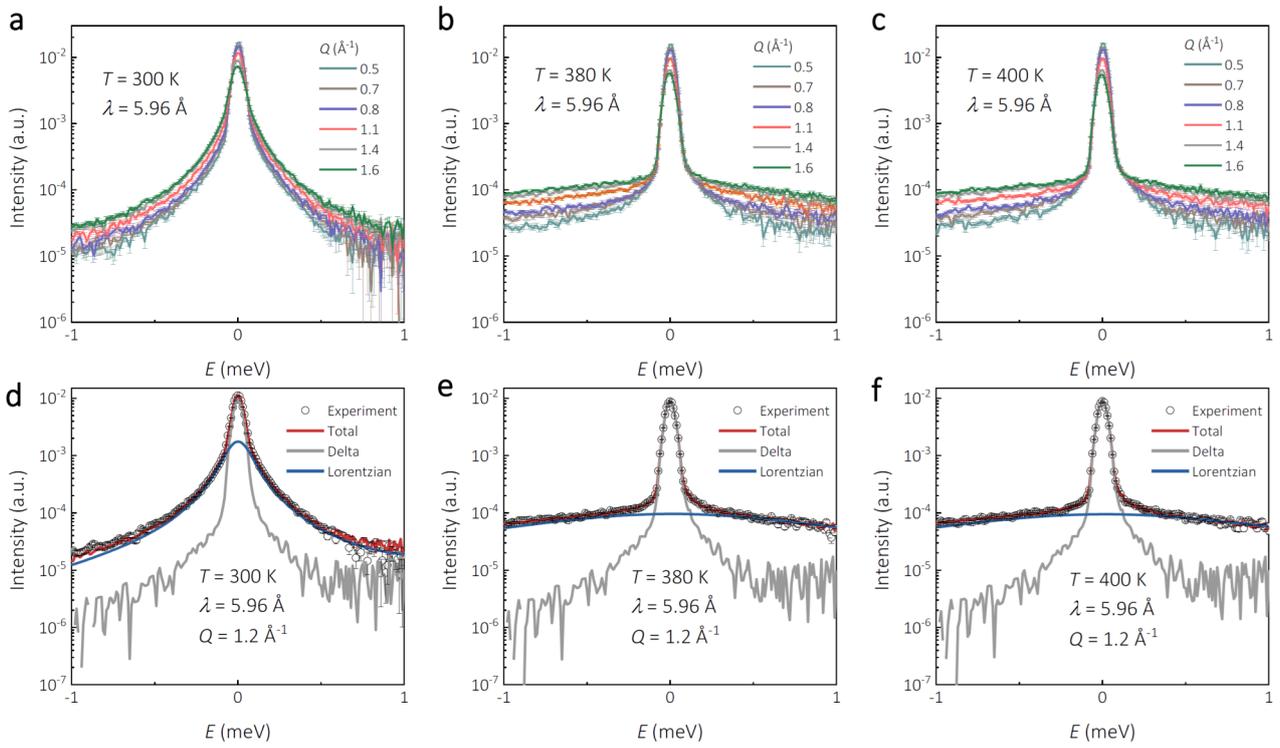

**Extended Data Fig. 5. QENS spectra and fitting. a**, **b** and **c**, Selected $Q$-point spectra at the temperature of 300, 380, and 400 K with the incident neutron wavelength of 5.96 Å. **d**, **e** and **f**, The fitting of QENS spectra of $Q = 1.2$ Å$^{-1}$ at the temperature of 300, 380, and 400 K with the incident neutron wavelength of 5.96 Å.

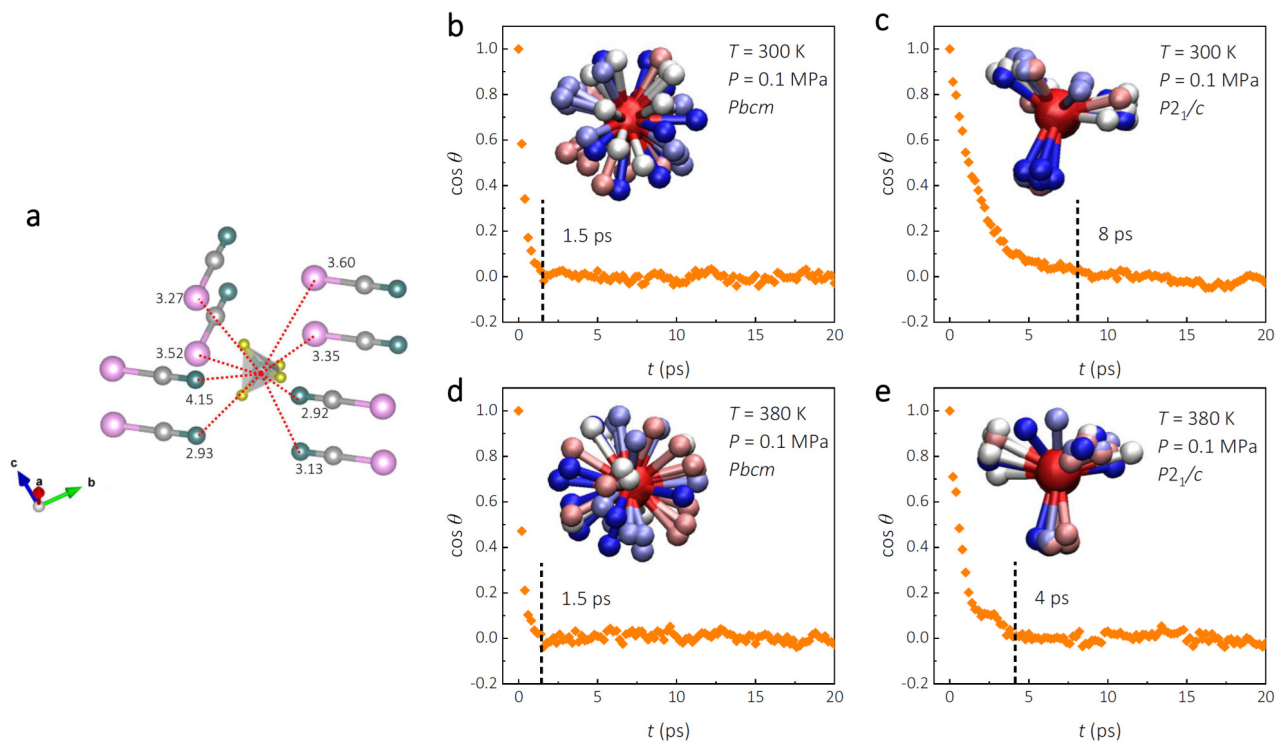

**Extended Data Fig. 6. Hydrogen bond distribution and relaxation times.** a. Hydrogen bond distribution of NH4SCN around $NH_4^+$. **b** and **c**, The cosine trajectory of N-H bonds in $NH_4^+$ cations in the *Pbcm* and *P2$_1$/c* phases at 300 K under pressures of 0.1 MPa. **d** and **e**, The cosine trajectory of N-H bonds in $NH_4^+$ cations in the *Pbcm* and *P2$_1$/c* phases at 380 K. The dashed line represents the estimated lifetime for the $NH_4^+$ rotation to be de-coherent from its initial orientation.

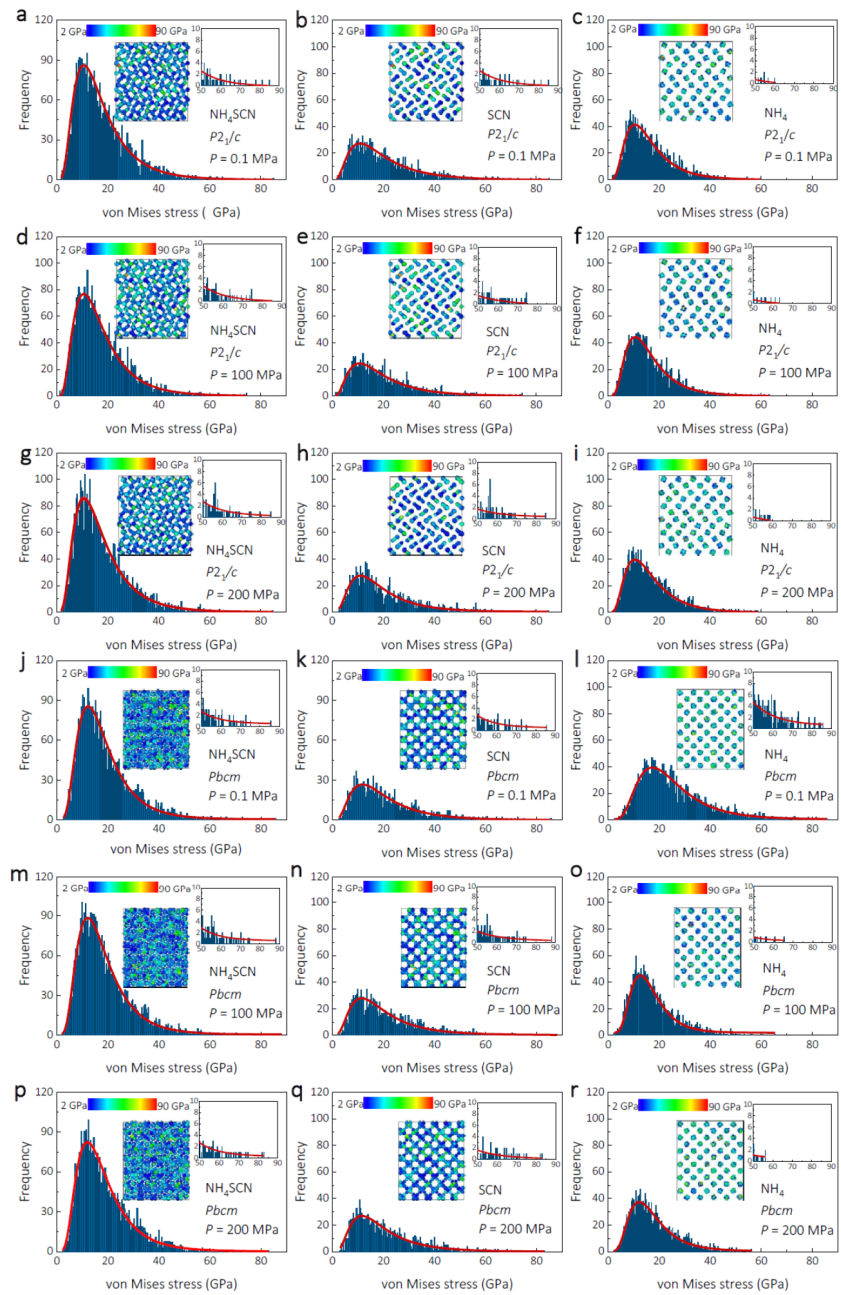

**Extended Data Fig. 7. Localization of the atomic von Mises stress on SCN⁻ in response to pressure.** a-i, Decomposed contributions from $NH_4^+$ and $SCN^-$ to the atomic stress of $NH_4SCN$ at $P2_1/c$ phase under different pressures (0.1, 100, and 200 MPa). j-r, Decomposed contributions from $NH_4^+$ and $SCN^-$ to the atomic stress on $NH_4SCN$ at *Pbcm* phase under different pressures (0.1, 100, and 200 MPa). Insets on the top center of each plot are colored by the local von Mises stresses. Insets on the top right of each plot are the probability distributions for von Mises stresses higher than 50 GPa, which are the major sources of stress localization.

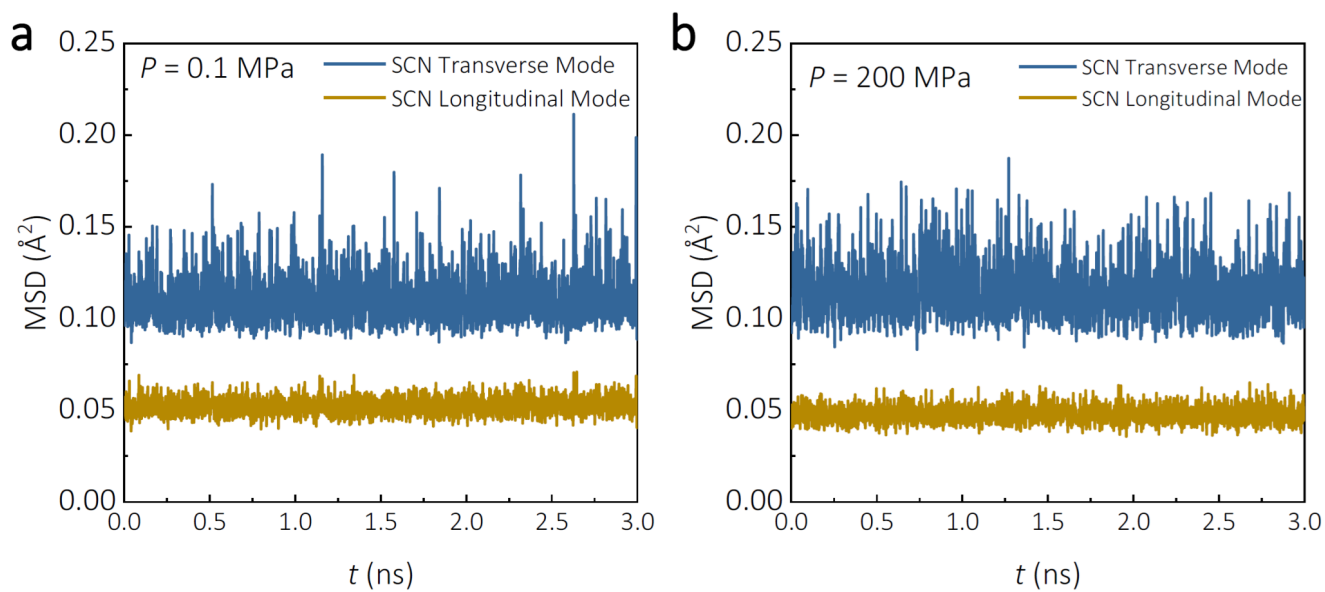

**Extended Data Fig. 8. Transverse and longitudinal vibrational modes of SCN⁻.** The time-evolution of the total mean square displacement (MSD) of SCN⁻ at 300 K and under 0.1 MPa (**a**) and 200 MPa (**b**) for the $P2_1/c$ structure, with decomposed contributions from transverse and longitudinal vibrational modes of SCN⁻.